\documentclass{ieeeaccess}

\usepackage{diagbox}
\usepackage{pict2e}
\usepackage{array}

\usepackage{amsmath}

\newtheorem{definition}{Definition}

\renewcommand\arraystretch{1.5}

\usepackage{cite}
\usepackage{amsmath,amssymb,amsfonts}
\usepackage{algorithm}
\usepackage{algpseudocode}
\usepackage{graphicx}
\usepackage{subfigure}
\usepackage{textcomp}
\usepackage{caption2}

\renewcommand{\algorithmicrequire}{ \textbf{Input:}}
\renewcommand{\algorithmicensure}{ \textbf{Output:}}
\def\BibTeX{{\rm B\kern-.05em{\sc i\kern-.025em b}\kern-.08em
    T\kern-.1667em\lower.7ex\hbox{E}\kern-.125emX}}
\begin{document}
\history{Date of publication xxxx 00, 0000, date of current version xxxx 00, 0000.}
\doi{10.1109/ACCESS.2017.DOI}

\title{SVM-Lattice: A Recognition \& Evaluation Frame for Double-peaked Profiles}
\author{\uppercase{Haifeng Yang\textsuperscript{1,2}},
\uppercase{Caixia Qu\textsuperscript{1}}, \uppercase{Jianghui Cai\textsuperscript{1,2}*}, \uppercase{Sulan Zhang\textsuperscript{1}} and \uppercase{Xujun Zhao\textsuperscript{1}}
\address{{[1]School of Computer Science and Technology, Taiyuan University of Science and Technology, Taiyuan 030024, China}
\address{[2]Shanxi Key Laboratory of Advanced Control and Equipment intelligence, Taiyuan 030024, China}
\tfootnote{The work is supported by the National Natural Science Foundation of China (Grant Nos. U1931209, U1731126) and Shanxi Province Key Research and Development Program (Grant Nos. 201903D121116, 201803D121059).}}}

\corresp{* Corresponding authors: Jianghui Cai (e-mail: jianghui@tyust.edu.cn)}

\begin{abstract}
In big data era, the special data with rare characteristics may be of great significations. However, it is very difficult to automatically search these samples from the massive and high-dimensional datasets and systematically evaluate them. 
The DoPS, our previous work\cite{qu2019dops}, provided a search method of rare spectra with double-peaked profiles from massive and high-dimensional data of LAMOST survey. The identification of the results is mainly depended on visually inspection by astronomers. In this paper, as a follow-up study, a new lattice structure named SVM-Lattice is designed based on SVM(Support Vector Machine) and FCL(Formal Concept Lattice) and particularly applied in the recognition and evaluation of rare spectra with double-peaked profiles.
First, each node in the SVM-Lattice structure contains two components: the intents are defined by the support vectors trained by the spectral samples with the specific characteristics, and the relevant extents are all the positive samples classified by the support vectors.
The hyperplanes can be extracted from every lattice node and used as classifiers to search targets by categories. 
A generalization and specialization relationship is expressed between the layers, and higher layers indicate higher confidence of targets. 
Then, including a SVM-Lattice building algorithm, a pruning algorithm based on association rules, and an evaluation algorithm, the supporting algorithms are provided and analysed.
Finally, for the recognition and evaluation of spectra with double-peaked profiles, several data sets from LAMOST survey are used as experimental dataset. The results exhibit good consistency with traditional methods, more detailed and accurate evaluations of classification results, and higher searching efficiency than other similar methods.
\end{abstract}


\titlepgskip=-15pt

\maketitle

\section{Introduction}
\label{Introduction}
\IEEEPARstart{I}n the context of massive and high-dimensional data, the research regarding special data with particular characteristics is becoming increasingly difficult\cite{saini2019data}. Searching of data with rare characteristics is a typical example proving the important of detecting and finding meaningful data from abundant special data. Some current search methods fail to mine these data with rare characteristics due to the properties of the data. The attributes presented as the rare characteristics are always considered together in detection works. The limitation noted above is one of the most influential factors causing the difficult classification of these data. 
In the previous works, a recognition method DoPS\cite{qu2019dops} is proposed based on relevant subspace and SVM for double-peaked profiles. In the DoPS, a characteristics subspace is extracted using relevant subspace mining algorithm. And the classification model is built using the support vectors trained from the labelled templates as thresholds. 
It is suitable for recognition of the double-peaked profiles. However, it can not evaluate the classification result. The relationships and diversity among the characteristics are not fully considered. An evaluation grid is built in this paper based on FCL to address above problem. As an effective data analysis tool, FCL has complexity and complete layer relationship. The relationship between layers conforms the characteristic connection of DoPS.
Thus, it is necessary to build a grid of the recognition and evaluation for the double-peaked profiles.
In this paper, we propose a recognition and evaluation method named SVM-Lattice for classification of data with rare characteristics, based on the Support Vector Machine and Formal Concept Lattice. The proposed method addresses the problem of classification when detecting special data with rare characteristics. Meanwhile, this approach reduces the negative influences between the attributes that are not necessarily in the characteristic data. In addition, we offer an effect evaluation method for the classification of unknown data.

Some symbols are included in this paper, which are listed in the following table to give a clear view. The meaning of symbols are in the Table \ref{tab_symbol}.
\begin{table}[htb]
\centering
\setlength{\abovecaptionskip}{0pt}
\setlength{\belowcaptionskip}{10pt}
\setlength{\tabcolsep}{4mm}
\caption{Symbol Summary}
\label{tab_symbol}
\begin{tabular}{ll}
\hline
Symbol&Meaning\\
\hline
SVM&Support Vector Machine\\
FCL&Formal Concept Lattice\\
K=(O, F, I)&formal concept\\
O&object set\\
F&feature set\\
I&binary relationship between O and F\\
o&object\\
f&feature\\
HP&hyperplane of DoPS\\
PS&extent of SVM-Lattice\\
SV&intent of SVM-Lattice\\
n1,n2,P1,P2,Q1,Q2&nodes of SVM-Lattice\\
S&support threshold\\
C&confidence threshold\\ 
Support(X)&support of item X\\
confidence(X$\Rightarrow$Y)&confidence of rule X$\Rightarrow$Y\\
pre($o_i$,c)&prediction of o on ith node with confidence c\\
\hline
\end{tabular}
\end{table}

\subsection{Motivations}
The recognition and evaluation method addressed of this study are mainly motivated by the following observations:

$\bullet$ Now, the identifications of the rare samples are largely depended on manual inspection by the astronomy experts. The complication and diversion of astronomical data improve the difficulty of automatically recognition and evaluations.

$\bullet$ For the spectra with double-peaked profiles, very few labelled samples are available relative to the astronomical big data, and the characteristics are very sparse and diverse relative to the high-dimension of the spectral data. 

$\bullet$ DoPS\cite{qu2019dops} can provide a useful hyperplane as classifier and the distance between the object and hyperplane can be regarded as its confidence. While, it does not fully consider the relationships and diversity among the characteristics. So, it would be meaningful to build a hyperplane grid according to the different characteristics. 

In the DoPS, a characteristic subspace is extracted using relevant subspace mining algorithm, including H$\alpha$, [OIII]$\lambda\lambda$4959, 5007, H$\beta$, [NII]$\lambda\lambda$6548, 6584, [SII]$\lambda\lambda$6717, 6731. And the characteristics set is analysed and grouped into 3 subsets according to the correlations among the characteristics based on the frequent patterns and rough set theory. Thus, the double-peaked profiles search algorithm is proposed by using the support vectors trained from the labelled samples as thresholds.

$\bullet$ FCL is always a useful tool of characteristics analysis, however, it is not suitable for solving the above problems directly.

{\bfseries Motivation 1:}
Samples with rare characteristics account for a small proportion of the massive data, which is an obvious difficulty for detection of these data from the large dataset. In addition to the low abundance, high dimensionality is another property of these data. There is some useless information in all dimensions due to the low quality of the data. However, the useful characteristics extracted from dimensions are various, including the profiles, qualities, and so on. Thus, a search method for the above mentioned samples is an important work for researches.  

{\bfseries Motivation 2:}
The spectra with double-peaked profiles found by researchers are total 345 from LAMOST and SDSS, which are  rare samples in the big datasets. And the characteristics are sparse and diverse relative to the dimensions in a spectrum.
The existing searching methods are applied to find some required data, including classification methods\cite{han2018new}, clustering algorithms\cite{li2019novel}, outlier detection algorithms\cite{zhao2020abnormal}, association rules mining\cite{sun2019incremental}, etc. These algorithms exhibit good performance in various fields, including image classification\cite{han2018new}, spectral clustering\cite{wang2018multiview}, credit card theft\cite{mishra2019comparative}, and so on. However, a few dedicated methods are used to detect the special data with above characteristics. Researching a method for detecting the rare samples in big data is necessary. Therefore, an effective model for detecting the rare samples from big data should be considered as the main work in this paper.

{\bfseries Motivation 3:}
A classification method DoPS\cite{qu2019dops} based on the SVM is a useful classifier, which serves as a hyperplane trained by the known samples with double-peaked profiles. A DoPS is obtained by the characteristics from feature extraction, as a threshold of the classification method. Before training the DoPS, the characteristics of double-peaked profiles are extracted by the relevant subspace. Thus, the eight characteristics are obtained to be used in the classifier of double-peaked profiles. 

{\bfseries Motivation 4:}
The formal concept lattice is build to discover the relationships between different transactions based on formal context.
It is always a useful tool of characteristics analysis, however, it is not suitable for solving the above problems directly. Thus, the combination of the DoPS and FCL in this paper is a meaningful method for the different characteristic combinations.

\subsection{Contributions}
For the detection of special data with particular characteristics, the properties of our data are so rare and varied that the search of these data is more difficult than searching for other balanced data. The support vector machine is used to be a threshold for nodes in the SVM-Lattice, which is a classifier for detecting our data from massive and high-dimensional datasets. Meanwhile, the lattice also can serve as an evaluation system for the classification results. The contributions in this paper are summarized as follows:

$\bullet$ A new lattice structure named SVM-Lattice is designed and theoretically analysed based on SVM $\&$ FCL.

$\bullet$ A SVM-Lattice building algorithm, a pruning algorithm and an evaluation algorithm are proposed.

$\bullet$ The proposed algorithms and techniques are integrated in SVM-Lattice, which is particularly tested on the recognition and evaluation of double-profile spectra using several datasets from the LAMOST survey.

\subsection{Roadmap}
The remaining content in this paper is arranged as follows: Sec.~\ref{sec:Related works} introduces the related works about classification, concept lattice, and SVM. In the Sec.~\ref{pri_method}, the primary knowledge and proposed method of classification are given. The methodology of the SVM-Lattice in this paper is introduced in Secs.~\ref{method} and \ref{description}. 
The experiment and analyses are shown in the Sec.~\ref{Experiments Analysis} from different perspectives. The summary of this paper is in the Sec.~\ref{Summary}. 
Finally, the acknowledgement of this paper including data resource and projects is shown in the Sec.~\ref{Acknowledgement}.

\section{Related works}
\label{sec:Related works}
The double-peaked profiles are meaningful for the research of the galaxies even universe. The search of the objects from massive celestial data is an important work. So the methods of the search are needed to be developed, including classification, clustering, and other mathematical methods. The previous works of double-peaked profiles dedicated to the specific objects, such as the reference \cite{kharb2019curved}, \cite{doan2020improved}, \cite{remillard2020transition}, etc. The machine learning based on parallelized hadoop is used to identify the double-peaked profiles on $H\alpha$ line in the LAMOST\cite{vskoda2016identification}. In the SDSS(Sloan Digital Sky Survey), a method involved a cross-correlation technique is proposed to detect the double-peaked or multi components profiles in galaxies\cite{garcia2013searching}.
However, the search of the double-peaked profiles data from massive spectra is a different work. The data with double-peaked profiles includes 345 spectra according to the previous search \cite{wang2018double,shi2014search}. Thus, developing an effective method for the search of double-peaked profiles is meaningful for the formation and evolution of universe, using the existing data conditions.

Classification algorithms are used to detect required data with some characteristics. In the past, different classification methods are applied widely in various fields. A combination of deep transfer learning CNN (Convolutional Neural Network) and web data augmentation based on feature presentation is proposed to address the problem of over-fitting on small datasets\cite{han2018new}. The CNN is a more popular method due to effective kernels in other application scenarios\cite{frid2018gan,zhu2018deformable,zhang2018diverse}. A machine learning algorithm of random forests is used to generate spatial and texture metrics for land-use mapping, which is achieving higher accuracy and superior coefficients compared to those of other approaches\cite{ruiz2018random}.
The rough set theory is developed with a similarity-based method to create a weight matrix scoring system, which produces a high accuracy with overall converge 67.47$\%$\cite{cekik2018new}. A pattern classification accuracy improvement method is proposed with a local quality matrix and is estimated based on the KNN (K-nearest neighbour) method. Outlier detection\cite{wahid2019distance} and the clustering method\cite{yang2020tad} are also regarded as classification tasks.
There are several other classification methods exhibiting high performance, including SVM (support vector machine)\cite{zhu2019achieving}, DNNs (deep neural networks)\cite{eykholt2018robust}, GAN (generative adversarial network), and so on. In addition, clustering methods are used for tasks of classification, including transfer clustering\cite{han2019learning}, deep clustering\cite{caron2018deep}, and so on.

As an unsupervised classification method, the SVM (support vector machine) algorithm is applied widely in different fields. A prediction of phage virion proteins (PVPs) based on support vector machine is trained with 136 optimal features, which displays superior performance\cite{manavalan2018pvp}. The kernel function is a key component in the construction of classification models, and some of the latest research efforts involve pursing the improvement of kernel functions to obtain better and more efficient training models\cite{hoang2018predicting,achirul2018comparison,zareapoor2018kernelized}. The mixture kernel function is employed to search for the optimal model parameters in the achievement of China’s carbon intensity target based on SVM\cite{zhu2019achieving}. The SVM algorithm also is used as a powerful classification tool for cancer genomic classification\cite{huang2018applications}. 
In addition to the single algorithm, a combination classification schema of SVM with other methods is applied to improve convective and stratiform classification, including random forest\cite{lazri2018combination}, logistic repression\cite{ing2019support}, neural network\cite{ye2019key}, etc. For the popularization of high-dimensional data, the SVM is employed to construct a classification model based on the feature selection or extraction method\cite{aljarah2018simultaneous,qu2019dops,cai2019feature}. By comparing other classifiers (random forest and KNN), SVM outperforms in terms of accuracy, with the least sensitivity to the training sample size\cite{thanh2018comparison}.

Formal concept analysis (FCA) or formal concept lattice (FCL), an effective tool for data analyses and knowledge extraction, was first proposed by German mathematician Wille in 1982\cite{wille2009restructuring}. The operations of the concept lattice itself are the topic of general research, such as the reduction of multi-adjoint concept lattice\cite{cornejo2018characterizing}. In addition to the above research, a hierarchical case-based classifier is proposed by introducing a concept lattice to hierarchically organize cases\cite{zhang2018hcbc}. In recent years, the FCL method has also been frequently combined with data mining methods\cite{braud2018generalization}. Fuzzy clustering and formal concept lattice are used to research public security\cite{de2018definition}. K-medoids clustering is used to compress an approximate concept lattice, which serves as robustness analysis of the proposed method\cite{li2016concept}. The formal concept lattice based on the support vector machine is also employed to address the problem of heuristic flame image feature selection and detection, which proves the enhanced recognition of features compared with manual work\cite{wen2016research}. Rough set theory and the concept lattice are combined to research the feature reduction, which provides a new reduction machine\cite{benitez2017attribute}. An algorithm for constructing a concept lattice based on cross data links can successfully cluster web pages\cite{zhang2018web}. Classification methods based on a concept lattice are proposed for different applications, including diabetes\cite{hao2019feature}, educational data mining\cite{beguvsic2018annotating}, spatial statistical services\cite{chen2018dynamic}, etc. However, there are few papers pertaining to searching for the combination of support vector machine and concept lattice for the classification of special data with particular characteristics.

It is seen that the concept lattice is a popular and effective method for data processing, which is fixed with respect to our data requirements. In viewing of the above methods for classification and concept lattice, there is little existing research about constructing a grid or lattice of hyper-plane based on the support vector machine. The notes in the grid comprise two parts, representing the association of objects and attributes. Thus, a method named the SVM-Lattice is proposed to provide a classifier for different characteristic combinations and obtain an efficient evaluation measurement method for classifiers.

\section{Primary Knowledge}
\label{pri_method}
The classification method DoPS is applied to train a classifier using the training sample. Meanwhile, the hyperplanes obtained from each combination of different features are used to construct a SVM-Lattice, which memorizes various hyperplanes trained by the dataset with the feature combinations. After obtaining a complete lattice, the meaningless nodes are reduced according to an association rule. Thus the final SVM-Lattice is a streamlined classifier with additional features.

\subsection{DoPS method}
Support Vector Machine, adapted for rare positive samples, can be used to generate a classifier for rare data with high dimensionality\cite{wadkar2020detecting}. Thus approach aims to find an optimal hyperplane with maximal margin between the separating hyperplane and other data\cite{jenkins2019general}. DoPS is proposed based on SVM to fit the spectra double-peaked profiles. Thus, the goal of DoPS is to find the minimum value of $2/\parallel \pmb{w}\parallel$, which is represented by the following equation:
\begin{equation}\label{equ1}
\begin{array}{l}
\min \frac{1}{2}{\left\| \pmb{w} \right\|^2} \\
s.t. {y_{i}(\pmb{w{x_i}} + b) \ge dic} \quad {i = 1,2,...,n}
\end{array}
\end{equation}

Where $\pmb{w}=\{w_1,w_2,...w_i,...,w_n\}$ is a row vector, while is a matrix composed of the weights of all support vectors, and $x_i$ is the ith object in the dataset, which is a column vector. In \eqref{equ1}, $n$ denotes the length of training dataset and $b$ is a constant computed by multiple iterations in training process. $dic$ represents a distance threshold between positive and negative support vectors, which is set manually.

To find the optimal support vectors, the Lagrange factors are introduced into the computation of $\pmb{w}$ and $b$. The solution of this process is a dual problem, which is converted to a new equation as follows:
\begin{equation}\label{equ2}
\begin{array}{l}
\mathop {\min }\limits_\alpha  \sum\limits_{i = 1}^n {{\alpha _i} - \frac{1}{2}} \sum\limits_{i = 1}^n {\sum\limits_{j = 1}^n {{\alpha _i}{\alpha _j}{y_i}{y_j}(\pmb{{x_i}\cdot{x_j}})} } \\
s.t.\left\{ {\begin{array}{*{20}{c}}
{\sum\limits_{i = 1}^n {{\alpha _i}{y_i} = 0} }&{}\\
{{\alpha _i} \ge 0}&{i = 1,2,...,n}
\end{array}} \right.
\end{array}
\end{equation}

Where $\alpha_i$ is the optimal Lagrange multiplier produced from the iterations. \eqref{equ2} is a linear objective function, but it is not suitable for the non-linear data. Thus, a factor is applied into the computation, the kernel function, which can map the input dataset into higher dimensional space. In the new space, the dataset is linearly separable using the kernel function. The Lagrange objective equation for adding kernel functions is as follows:
\begin{equation}\label{equ3}
\begin{array}{*{20}{l}}
\mathop {\min }\limits_\alpha  \sum\limits_{i = 1}^n {{\alpha _i}  - \frac{1}{2}} \sum\limits_{i = 1}^n {\sum\limits_{j = 1}^n {{\alpha _i}} {\alpha _j}{y_i}{y_j}K(\pmb{{x_i}},\pmb{{x_j}})}\\
{s.t.\left\{ {\begin{array}{*{20}{c}}
{\sum\limits_{i = 1}^n {{\alpha _i}{y_i} = 0} }&{}\\
{0 \le {\alpha _i} \le C}&{i = 1,2,...,n}
\end{array}} \right.}
\end{array}
\end{equation}

Where $K(\pmb{{x_i}},\pmb{{x_j}})$ is the kernel function selected before training process. As an important factor for the training result, the kernel function is considered as the one of most significant parameters. There are several types of kernel functions, such as the linear kernel, Gaussian kernel, and sigmoid kernel. Among them, the first is the easiest kernel, which requires the least running time. The Gaussian kernel is the most frequently used in the process of training the classification model. However, the increased running time is the one shortcoming when the Gaussian kernel function is applied in the SVM. 
By comparing the Gaussian and linear function, it is found that the same classification result is obtained with the two functions. Thus, the linear kernel function is applied in DoPS.

In \eqref{equ3}, $C$ is a parameter named penalty factor, which is the tolerance for the error. $C$ is introduced into the objective function because the hard margin does not tolerate the classification error. The soft margin with $C$ has greater tolerance for misclassified objects.
The higher $C$ is, the lower the accepted tolerance and the easier over-fitting becomes. The smaller $C$ is, the easier fitting becomes. If $C$ is too large or small, the generalization ability will become worse.

After joining the kernel function and soft margin, the prediction classifier trained by training data is constrained as follows:
\begin{equation}\label{equ4}
f(x) = \sum\limits_{i = 1}^L {\sum\limits_{j = 1}^m {{P_i}{w_{ij}}{x_{ij}}} }
\end{equation}

Where $\pmb{x_i}$ is a data object to be classified, which is designed by the value of $f(x)$. $P_{i}$ is the probability of the $ith$ feature in feature subspace of DoPS. If $f(x)$ is close to positive line, then the object is considered to be a positive sample. It will be designed as a negative sample when its predicted value is close to the negative line.

In the classification of DoPS, the positive and negative samples are divided into areas aside the hyperplane. In other words, the object above the positive line (under the negative line) is considered to be in the positive class (negative class). The remaining objects between two lines represent the fuzzy data, which can be determined by other effective methods, such as the sort algorithm and fuzzy set algorithm.

\subsection{Formal Concept Lattice}
The concept lattice was first proposed by Wille in 1982 to provide an effective tool for data analyses and knowledge extraction\cite{kapoor2020crime}. There are some obvious properties in the total concept lattice, including completeness, intuitiveness, simplicity and so on. Thus, this approach is more popular in the data processing field due to its non-negligible advantages.

The construction of the concept lattice is based on a form background\cite{qin2020local}, which is a triple of $K=(O, F, I)$. Giving a formal description for K is the main target of concept lattice. The definition of the formal background is as follows:
\begin{definition}\label{def1}
There are two sets $O, F$ and a binary relationship $I$ in a given form background $K=(O, F, I)$. $O$ and $F$ are data set and feature set respectively, and $I\subseteq O\times F$ is a binary relationship of object set and feature set. The object $o\in O$ is existed with feature $f\in F$ if $(o,f)\subseteq I$ or $'oIf'$.
\end{definition}

In Definition \ref{def1}, relationship $I$ is composed of an object and feature. The formal concept lattice is constructed according to the background in Definition \ref{def1}. 

A number of nodes is included in the concept lattice, representing the component elements in the total lattice structure. The introduction of each node is seen in the follow definition.
\begin{definition}\label{def2}
Each node is represented by a formal concept $J$, which is an order ($W, N)$. The $W\in O$ and $N\in F$ are extent and intent respectively. $W$ is a largest subset of objects with intent $N$, meanwhile, $N$ is a subset of common feature including $W$.
\end{definition}

The formal concept in Definition \ref{def2} is used to construct a formal concept lattice, which is defined in Definition \ref{def3}.
\begin{definition}\label{def3}
In a formal background $K=(O, F, I)$, the relationship between any two formal concepts is a partial order relationship, which is represented by $(A_1, B_1)\leq (A_2, B_2)\Longleftrightarrow A_1\subseteq A_2 \Longleftrightarrow B_2\subseteq B_1$. The all formal concepts and their partial order relationships are made up a formal concept lattice named $<L(O, F, I), \leq>$.
\end{definition}

A complete lattice is obtained according to the formal context, including the formal concept and their order relationship. In the concept lattice, the parent nodes and child nodes are mutually corresponding. The definition of these nodes are in Definition \ref{def4}.
\begin{definition}\label{def4}
Two different nodes $n1(A_1, B_1), n2(A_2, B_2)$, $n_1 < n_2\Longleftrightarrow B_2\subset B_1 \Longleftrightarrow A_1\subset A_2$. If there are no other nodes $n_3$ with $n_1 <n_3 < n_2$, $n_2$ is the parent node(parent concept) of $n_1$, meanwhile, $n_1$ is the child node(child concept) of $n_2$.
\end{definition}

There are a number of levels in a concept lattice according to the feature combination. Any node in a concept lattice has at least one related node, which is the parent or child node. For the child nodes in the next level, the intents (features) are the intersections of the intents in parent nodes. In a word, the nodes in the concept lattice are interpretable and completed.

\section{SVM-Lattice Frame}\label{method}
\subsection{Hyperplane Lattice}
The extent and intent of each node in the concept lattice indicate the direct relationship according to the formal context. The intent is the common feature of the extent, which can be viewed in the provided formal context. A new mapping is introduced to express the relationship between extent and intent in our proposed method. Each node of the lattice is an independent concept using every hyperplane obtained by DoPS training dataset. Thus, the hyperplanes using datasets with different feature combinations will constitute a new lattice called the SVM-Lattice.

\subsubsection{Completed SVM-Lattice}
In a given dataset with high dimensions, the rare characteristics appear in different locations. The characteristics that exist in the data are combined randomly to make up different training datasets, including positive and negative samples. Different training datasets will be used for constructing classification model due to the uncertainty of characteristic locations. Thus, the certified known sample with characteristics will be labelled to show which exist in an object. The labelled sample is used to combine the positive dataset with characteristic combination. After obtaining the training dataset, a classification model with one characteristic combination is trained by the DoPS. This means that a hyperplane will be obtained using the training dataset. 

The definition of the new extent is shown in Definition \ref{def_ex}, which is determined to represent the positive sample upper hyperplane.

\begin{definition}\label{def_ex}
The hyperplane $HP$ is obtained from the DoPS using the training dataset $TR$. The testing dataset is used to test the effectiveness of $HP$. The objects upper $HP$ are regarded as positive sample to be the extent of new concept, abbreviated $PS$.
\end{definition}

Another component of the formal concept is intent, the support vectors in formal context. The definition of the new intent is shown in the follow Definition \ref{def_in}.

\begin{definition}\label{def_in}
The SVM-Lattice is a new lattice obtained by SVM due to the redefined intent. Different support vectors (SV) are calculated by training processing, which are regarded as the new labels on characteristic combinations. Each combination of support vectors is the intent of each node in SVM-Lattice on different characteristic combinations. 
\end{definition}

The obtained hyperplane with characteristic combinations is a boundary line between positive and negative data. The predicted positive sample above hyperplane is regarded as the extent of a node(formal concept), meanwhile the intent of node is the support vectors on characteristic combinations. The redefined formal concept is called the hyperplane concept, whose structure is shown as Definition \ref{def_pseudo}.

\begin{definition}\label{def_pseudo}
Two nodes $n1=(N1, N2), n2=(M1, M2)$ are sequences with Definition \ref{def3} to make up a SVM-Lattice $<L(PS, SV, HP),\leq>$ according to Definition \ref{def4}. The $HP$ is denoted a relation between $PS$ and $SV$, which means $PS$ is a dataset selected by relation $HP$ on intent $SV$.
\end{definition}

\subsubsection{Pruning the SVM-Lattice}
All of the nodes on characteristic combinations are included in a completed SVM-Lattice. However, users devote less attention to some nodes or useless information due to the actual situation. The hyperplane concepts regarded as reduced nodes should be removed from the complete SVM-Lattice. The association rule algorithm is applied as the pruning method for our SVM-Lattice. The association rule is a machine learning method based on rules, and it is used to find the implicit relationship ($A\Rightarrow B$) between transactions from a sample. The method can find the relationship between basket data and other data types, including medical diagnosis\cite{sharma2019concept}, web data\cite{de2019multirelation}, risk assessment\cite{saranyadevi2019road}, and so on.

Two steps are included in the process of the association rule: finding frequent items and association rules\cite{krishna2019data}. The first step is a comparison between items and support threshold $S$. All items obtained by scanning the database are compare with $S$ to represent frequent items, which are used for finding associations rules in the next step. Different rules produced according to the frequent items are shown with values of confidence. The confidence threshold $C$ is set to select the most reliable rules with higher $C$. 
The values of $S$ and $C$ are set before beginning association rules, being compared with the support confidence of items. The computations of the two values are as follows:
\begin{equation}
\label{equ5}
\begin{split}
&Support(X)=\frac{\#X}{n}\\
&Confidence(X\Rightarrow Y)=\frac{Support(X\cup Y)}{Support(X)}
\end{split}
\end{equation}

Where $Support(X)$ is the support of item $X$ and the $Confidence(X\Rightarrow Y)$ is the confidence of rule $X\Rightarrow Y$. In addition, the $\#X$ and $n$ are the frequency of item $X$ and the length of the database.

The detailed steps of finding association rules are shown in Algorithm \ref{alg_apri}.

\begin{algorithm}[htb]
\caption{Pruning Method Based on Association Rule}
\label{alg_apri}
\begin{algorithmic}[1]
\Require formal context FC, support threshold S and confidence threshold C.
\Ensure Reduced intents
\State Scanning database CL to find the 1-items;
\State Computing the frequency of 1-items to get the frequent 1-items by comparing with S;
\For {1<k<length of FC}
\State Connecting the (k-1)-items to produce the k-items;
\State Pruning k-item using the prior knowledge to remove the useless item from k-items;
\State Obtaining the frequent k-items by comparing the support with S;
\EndFor
\State Producing the rules and computing their confidence according to \eqref{equ5};
\State Recording the confidence of subsection of each rules;
\State Selecting the subsections with higher confidence as the final intents;
\State Return reduced intents.
\end{algorithmic}
\end{algorithm}

The pruning algorithm using the association rule method is a reduction of the completed SVM-Lattice. The scanning process included in the reduction occurs for each determination of the support of the items. The repeated scans require more memory and time; however, this can be ignored in this paper due to the smaller size of the formal context. Thus, the Apriori method is used as an association rule for mining the rules.

\subsection{Natures of the SVM-Lattice}
The extent of concepts in the SVM-Lattice is mapped indirectly using support vector machine. 
Properties of formal concept lattice are considered to be the natures of the SVM-Lattice. In addition, the generalization and specialization of relationship between nodes in different layers should also be analysed from several angles. The natures of the SVM-Lattice are analysed in the following descriptions.

\subsubsection{Number of nodes}
The number of nodes and edges in SVM-Lattice are smaller than those in the formal concept lattice. 
Any intent in formal concept lattice is all the support vectors on feature combinations of the initial formal context. However, the intents in the SVM-Lattice are composed of the support vectors attracting users and not include those useless. Thus, the number of intents in the SVM-Lattice is smaller than that in the initial lattice. The edges in a lattice are related to nodes due to the binary relationship $I$ in the SVM-Lattice. Thus, the number of nodes and edges in the SVM-Lattice are smaller than those in the formal concept lattice.

\subsubsection{Complexity}
The complexity of the SVM-Lattice is lower than that of the formal concept lattice. 
The process of pruning for the formal concept lattice is included in the construction of SVM-Lattice, in which the number of nodes and edges is smaller than those in the formal concept lattice according to the first nature of the SVM-Lattice. Thus, the complexity of the SVM-Lattice is lower than that of the general lattice structure.

The relation between nodes and layers can serve as context to help detect hidden information. The finding of detection is an important step in classification of data with rare characteristics. Thus, the classification using the SVM-Lattice exhibits greater complexity than DoPS classification due to the process of finding the context.

\subsubsection{Completeness}

In a formal concept lattice $<L(O, F, I), \leq>$, the order sequence (concept) is completed with respect to the relationship $I$. A given order $P(O, F)$ is constrained by the completeness as follows:\\
\begin{equation}
O = \lbrace o\in O\vert \forall f\in F, oIf\rbrace
\end{equation}
\begin{equation}
F = \lbrace f\in F\vert \forall o\in O, OIF\rbrace
\end{equation}

A given node $P'(O', F)$ in the SVM-Lattice, corresponding to $P$, is also constrained by the above two conditions. $\forall o\in O(f\in F)$ is satisfied relationship $oIf$ if $f\in F(o\in O)$. $O'$ is a positive sample within the upper hyperplane from DoPS, and thus, it is regarded as a corresponding mapping as follows:\\
\begin{equation}
O' = \lbrace o'\in O'\vert \forall f\in F, o'If \rbrace
\end{equation}
\begin{equation}
F = \lbrace f\in F\vert \forall o'\in O', O'IF\rbrace
\end{equation}

Completeness is a basic property of our SVM-Lattice indicating that the sequence with the largest expansion will appear in the lattice structure. A node in the formal concept lattice is mapped into another lattice space by the DoPS. If a concept $P(E, I)$ is a directed map in the formal concept lattice, E and I meet the above conditions. Meanwhile, in the SVM-Lattice, node P's indirected mapping still satisfies completeness.

The nodes in the SVM-Lattice are the subset of completed lattice due to the pruning process. It is necessary to remove the nodes with lower confidence, which are useless for data searching. The removed nodes are regarded as meaningless points by combining expert knowledge. Thus, even with pruning, the completeness of the SVM-Lattice still exists.

\subsubsection{Relation between nodes}
Nodes in the same layers are relatively independent, while those on different layers are connected and related.
On the same layers, the intents of nodes are support vectors on different characteristic combinations selected as meaningful information for users. These nodes from at least one same father node include crossed characteristics, which are the intersections of their father's intents. However, nodes in the same layer from different fathers are rarely related. For the nodes of different layers, on the one hand, the node is a child of another node on the upper layer. On the other hand, their father nodes are not the same nodes and are not related. 

\begin{figure}[htb]
\centering
\subfigure[no-crossed of P1 and P2]{\label{fig3:a} 
\includegraphics[scale=0.35]{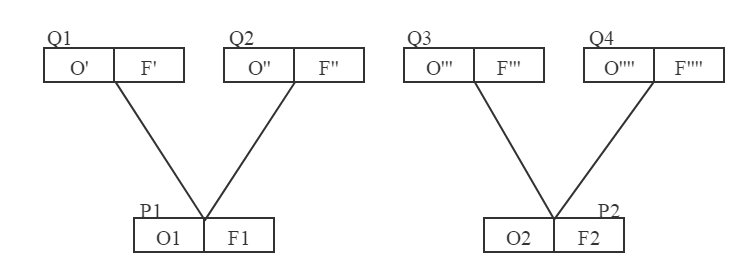}} \hspace{2in} 
\subfigure[crossed P1 and P2]{\label{fig3:b} 
\includegraphics[scale=0.35]{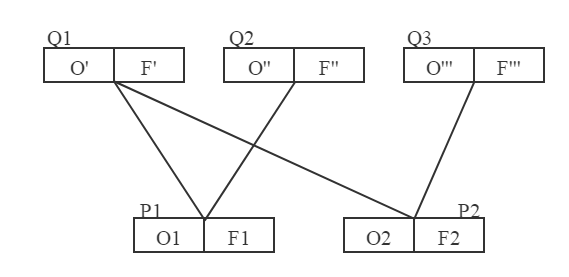}} 
\caption{Relationship of Position Between Nodes in the Same Layer}
\label{fig3}
\end{figure}

For a brief addition to the above description, it is assumed that two nodes $P1(O1, F1)$ and $P2(O2, F2)$ exist. If $P1$ and $P2$ are on the same layer, the possible relationship of the two nodes is shown in Figure \ref{fig3}. The top sub-figure in Figure \ref{fig3} represents a non-crossed relationship in which $P1$ and $P2$ are from different father nodes. It is observed from Figure \ref{fig3:a} that $P1$ and $P2$ are relatively independent. The intents of $P1$ and $P2$ are with $F1\cap F2 = \emptyset$. The crossed relationship between $P1$ and $P2$ exists in the bottom sub-figure, in which $Q1$ is the crossed parent node of $P1$ and $P2$. There is at least one common characteristic of $P1$ and $P2$ with $F1 \cap F2 = F'$, meaning that $P1$ and $P2$ are characteristic expansions based on $Q1$. There is more intent and less extent in $P1$ and $P2$ due to the characteristic expansion. In addition, the extents of two nodes are related by node $Q1$ with $O'\cap O'' = O1 $ and $O'\cap O''' = O2$. This observation indicates that the $O1 \sqsubseteq O'$ and $O2 \sqsubseteq O'$, and it is possible with $O1 \cap  O2 = \emptyset$.

\begin{figure}[htbp]
\centering
\setlength{\abovecaptionskip}{10pt}
\setlength{\belowcaptionskip}{5pt}
\subfigure[no-crossed of P1 and P2]{\label{fig4:a}
\includegraphics[scale=0.35]{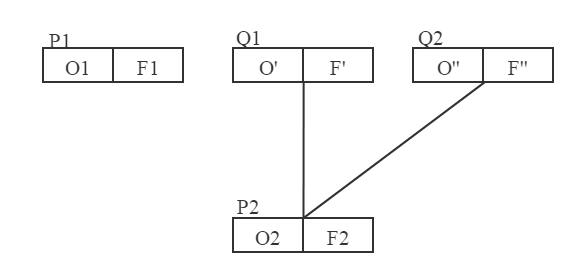}} 
\vspace{-0in}
\subfigure[crossed P1 and P2]{\label{fig4:b}
\includegraphics[scale=0.35]{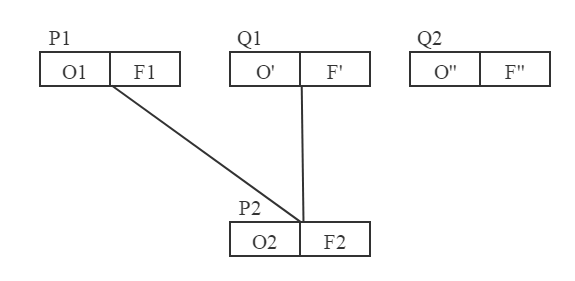}}
\caption{Relationship of Position Between Nodes in Upper and Lower Layer}
\label{fig4}
\end{figure}

The upper and lower relationships are shown in Figure \ref{fig4}, in addition to the same layers of $P1$ and $P2$.
The top plot is a non-relationship between $P1$ and $P2$, which are from upper layer and lower layer, respectively. The intent of $P2$ is composed of $Q1$ and $Q2$ in \ref{fig4:a}, while $P1$ does not participate in the formation of $P2$. In contrast, $P2$ is the child node of $P1$ in \ref{fig4:b}. The intent of $P1$ is the subsection of $P2$ according to the construction process of the lattice. Thus, $F2$ is a combination of $F1$ and other intents and is named $F2\subseteq F1$. $P2$ is denoted as a hyperplane node including characteristic combination $F2$ and positive sample $O2$ selected by DoPS. The two hyperplanes of $P1$ and $P2$ are trained by their training samples, which are on characteristic combination. $P1$ and $P2$ are related with their characteristics due to the crossed intents. For the extents of these nodes, $O2$ is a dataset constrained by $F1$ based on $F'$. Thus, $O2$ is the intersection of $P1$ and $Q1$ with $O1\cap O' = O2$. The size of $O2$ is smaller than $O1$ due to the greater feature constraint. Thus the more characteristics there are, the smaller training objects are.

\subsubsection{Relation between layers}
The generalization and specialization of relationship between different layers in the SVM-Lattice are analysed according to the searching path and distance between layers. The search of best intent for a given object is initiated from the bottom node of the SVM-Lattice. Each node in the SVM-Lattice is shown with a confidence according to the association rules algorithm. In the searching process, the distance between the upper and lower layers is computed to determine whether the search ends. 

The prediction of object $o$ on the $ith$ node in the lower layer is obtained by the DoPS with $pre(o_i, c)$. $o_i$ and $c$ are predictions of $o$ on the $ith$ node and confidence, respectively. All predictions of the lower layer are included in a prediction set with $pre = \lbrace pre(o_i, c)\rbrace$. Node $n$ in the lower layer is obtained by $min(pre)$ with higher confidence. Node $m$ with minimal prediction in the upper layer is also obtained by using the same process as that of the lower layer. The distance between the lower and upper layer is computed by $dis(m, n)$, which is the distance between two nodes $m$ and $n$. The search process can be ended according to the distance between two layers. It is assumed that, for three layers $L1, L2, L3$ from bottom to top, the search will be ended if $dis(L1, L2) < dis(L2, L3)$. There is no need to continue the calculation of the remaining layers, which reduces the search time and complexity. 

\subsubsection{Usage}
The SVM-Lattice can serve as an evaluation of classification results for special data with rare characteristics according to hyperplane on every characteristic combination. 
After forming a SVM-Lattice, every node in the lattice denotes a hyperplane of its characteristic combinations. The accuracy of the classifier for one intent is added into the node as additional information to offer an evaluation of classification with respect to this characteristic combination. All nodes in the SVM-Lattice are used to build a grid to provide a reference for users. In the grid, every point is a hyperplane representing a classification result for a characteristic intent. A given sample can be compared with the point in the grid to help determine which intent is more valid. The larger the distance is between the computed hyperplane and existed hyperplane, the less believable is the classification of the characteristic of this node. This evaluation system is meaningful for determining the best classification characteristic combination of special sample.

\section{Description of the SVM-Lattice}\label{description}
\subsection{Build of SVM-Lattice}
The introduced algorithm in this paper is referred to as the SVM-Lattice algorithm, which is utilized to build a classification lattice for special data with rare characteristics. This method is proposed to form a grid based on the DoPS and formal concept lattice. Each node in the lattice is denoted as a hyperplane by the training dataset, including positive samples and characteristic combination. The algorithm description is available in Algorithm \ref{alg_hy}.

\begin{algorithm}[htb]
\caption{SVM-Lattice Build Algorithm}
\label{alg_hy}
\begin{algorithmic}[1]
\Require Training sample D, characteristic combination F, characteristic matrix FM with 1 and 0
\Ensure A SVM-Lattice
\For {each characteristic f in FM}
\State select point with 1 on characteristic f;
\State add f to the characteristic combination F;
\EndFor
\For {each feature combination f in F}
\State select training sample on f from D;
\State select a type of kernel function K;
\State matrix the training data;
\State generate a matrix of kernel value KM;
\State obtain an initial matrix of Lagrange factors $\alpha$;
\State update the $\alpha$ according to function \eqref{def3};
\State produce the optimal $\alpha$ by iterations by DoPS;
\State produce the support vectors SV on f by DoPS;
\State filter out positive sample upper the hyperplane to be as extent on f;
\State ensure SV as the extent of this node;
\EndFor
\State remove out the reduced nodes from completed SVM-Lattice by association rule according to Algorithm \ref{alg_apri};
\State return a reduced SVM-Lattice.
\end{algorithmic}
\end{algorithm}

In Algorithm \ref{alg_hy}, a characteristic matrix is firstly produced by using the positive sample, including 1 and 0. If the double-peaked profiles appear in $ith$ data on $jth$ characteristic, the value of these positions in the characteristic matrix are set as 1. In contrast, the values are 0 for $ith$ data with respect to the $jth$ characteristic. The matrix is used to find all feature combinations for which the rare characteristics appear. On different characteristic combinations, there are various sample sizes for training the classifier. There are two parts in the algorithm: finding characteristic combinations and training the classifier on each characteristic combination. In the first process, the time complexity of constructing feature combination $F$ is $O(L)$ in which $L$ is the characteristic length of the characteristic matrix. The training process of SVM in the second part is the main component of constructing the SVM-Lattice. On each combination, the time complexity of the training part is $O(n\cdot d)$ with the linear kernel function, while $n$ and $d$ are the lengths of sample and characteristic. This parameter is $O(n^2\cdot d)$ when the kernel function is non-linear. The total time complexity of the entire process of constructing a SVM-Lattice is $O(n\cdot d\cdot FL)$, in which $FL$ denotes the maximum length of characteristic combinations. The special data with rare characteristics are extremely rare in massive data, so less time is required to construct a SVM-Lattice.

\subsection{Evaluation Process of the SVM-Lattice}
As an efficient classification evaluation method, the SVM-Lattice in this paper is constructed according to the formal context. The special data with rare characteristics are applied to build a lattice for the recognition and evaluation of classification. The data used for the classification should be analysed with respect to the dimensions, characteristics, and profiles, among others. Based on this type of data, a SVM-Lattice for the classification method is built from the formal context to begin the total process.
First, a formal context is given which includes the existence on each characteristic of characterised data, which is the known data certified by field experts. Second, the formal context is used to generate all nodes in the SVM-Lattice, including intent and extents by the DoPS. Third, the association rule algorithm is applied to find the redundant nodes, aiming to remove the nodes with lower confidence from the set of all nodes. Finally, an evaluation method of classification results is provided for the data with rare characteristics using the SVM-Lattice. 

Within a given dataset that must be assigned one characteristic combination, the shortest distance from the hyperplane of each node can be found. The distance between layers for given data is computed to judge the search for the best layer. The search of the best layer is started from the first layer, and it is ended when next layers are more distant. The search is interrupted when it is not necessary for the next nodes to be compared continuously. Therefore, all of the layers and nodes in the SVM-Lattice are denoted with meaningful information of the hyperplane by the DoPS.

The evaluation process of the SVM-Lattice is shown in Algorithm \ref{alg_process}, which is a detailed search process for a given object.

\begin{algorithm}[htb]
\caption{Evaluation Process of SVM-Lattice}
\label{alg_process}
\begin{algorithmic}[1]
\Require a SVM-Lattice, unknown data $D$, prediction value threshold $\sigma$
\Ensure evaluation information of $D$
\State Building an evaluation grid using the nodes in SVM-Lattice;
\For {each object $o$ in $D$}
\State Generating a new data $nd$ according to the intent of node $n$;
\State Computing the prediction value $pre(nd, n)$ on the intent of $n$;
\State Obtaining the confidence $c(n)$;
\If {$pre(nd, n) in [-\sigma, \sigma]$}
\State Jumping to the related node $rn$ in the next layer;
\State Repeating the previous two steps;
\If {$c(n)>c(rn)$}
\State Determining $o$ is between the node $n$ and $rn$;
\Else 
\State $o$ is on the node $n$;
\EndIf
\Else 
\State Jumping to the next node;
\State Repeating the step 2 to 11;
\EndIf 
\EndFor
\State Return the closed nodes with confidence of $D$.
\end{algorithmic}
\end{algorithm}

\subsection{Theoretical Analysis}
The two main tasks of the proposed SVM-Lattice in this paper are classification of special data with rare characteristics and evaluation for classification. The two parts are sequentially performed by the SVM-Lattice. For a given object that needs to be assigned, each node in the SVM-Lattice is computed from upper to lower layers. The node is denoted as a hyperplane by the DoPS, where the prediction value of the object is obtained. The distance between object and hyperplane is obtained by prediction, which is the basis for determining the closet node. In addition to the distance between the object and each node, the distance between nodes are calculated to measure the similarity of different nodes. Moreover, the differences between layers for a given object can be computed to find the nearest layer. 

The trained hyperplane of each node is obtained with a fixed confidence to judge the credibility of classification with respect to the intent. The association rules method is used to obtain the confidence according to operations on characteristic sets. The higher the confidence is, the more reliable the classification result is. The information of one node includes not only the classification components but also the evaluation system. After obtaining the prediction of each, the distances between nodes are used to find the nearest node on one layer. Therefore, the two aspects of classification and evaluation are interdependent. The classification process begins with the top node in the SVM-Lattice, while it is not ended at the lowest node. When the number of characteristics involved in the calculation increases, the accuracy and precision rate become higher than one the single characteristic combination. The build of the final SVM-Lattice is completed by combining all characteristic combinations, which costs more time due to the training process of the classification model. However, the SVM-Lattice will not be changed once it is built. The classification of unknown data simply traverses the SVM-Lattice. Overall, the classification based on the SVM-Lattice is more efficient and accurate.

\begin{figure}
\centering
\subfigure[build flowchart]{
\includegraphics[scale=0.3]{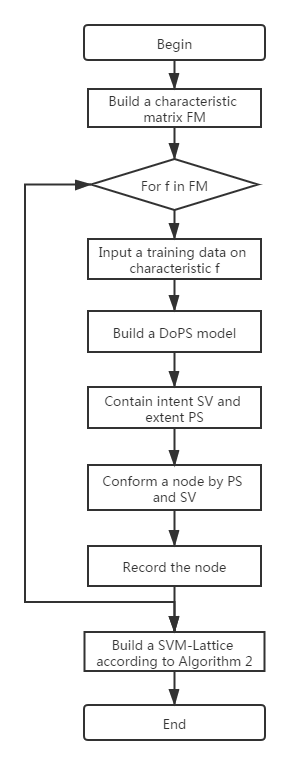}
\label{time_a}}
\subfigure[evaluation flowchart]{
\includegraphics[scale=0.3]{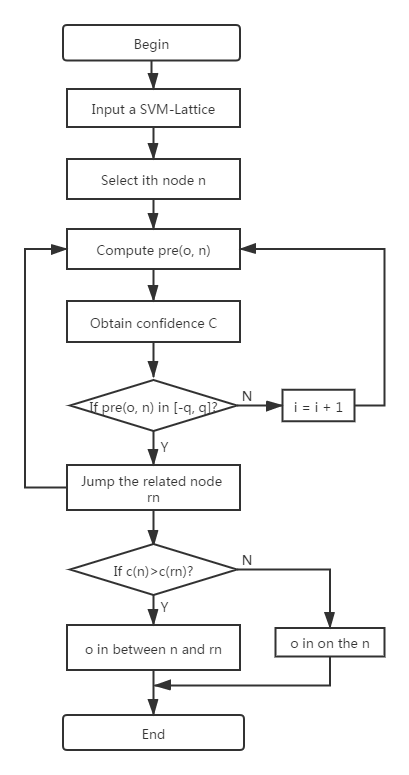}
\label{time_b}}
\caption{The recognition and evaluation flowchart}
\label{flowchart}
\end{figure}

The build and evaluation process is shown in Figure \ref{flowchart}. The SVM-Lattice is built with two roles, recognition and evaluation of the data with rare characteristics. In the recognition process, it is a classification in each node for a given data. The object is compared with each node from top to bottom until the best location is found. Each node denotes a DoPS classifier on one characteristic combination, which is a threshold of the classification. The confidence is given in each node to provide the evaluation of classification result for users. The object is matched on different characteristic combinations to adopt the characteristics of the object. The best location is on a node or one between nodes and layers. The final result of object is a location with a confidence, instead of a classification predict than other methods.

\begin{figure}[htp]
\centering
\includegraphics[scale=0.35]{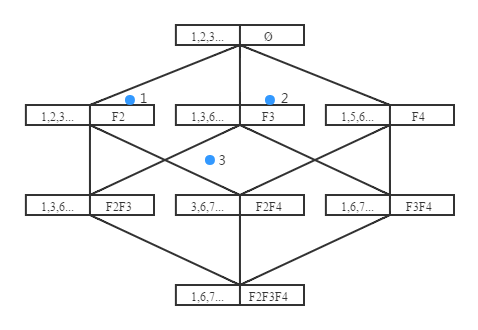}
\caption{A Search Process Example based on Three Characteristics. The first part in each node is the extent dataset, and the second one is the intent on characteristic F. The blue point is a given object, beginning with the top node. Labelled 1, 2, and 3 aside point are the sequence of the search process.}
\label{fig_theo}
\end{figure}

As an example, three characteristics in characteristics spectra, [OIII]$\lambda\lambda$4959,5007, H$\beta$, are selected from 8 characteristics to show the search process for a given object. The process of finding the best location is shown in Figure \ref{fig_theo}. The top node is ignored due to the empty intent. Therefore, the blue point is started at the first node, labelled 1. However, this is not the best node for the point. It then jumps to the second node, labelled 2, which is the first finding node after computing the prediction. Node 3 is selected to be compared with the last node. Finally, the point is located between two nodes, 1 and 2. The confidence of this position is given as a judgement of the search result.

\section{Experiments Analysis}
\label{Experiments Analysis}
\subsection{Data Preprocessing}
The spectral data with double-peaked emission lines, meaningful for researching the formation and evolution of galaxies, and even the universe, are used to test the availability of our method. It is significant to search the double-peaked emission lines spectra for finding some rare celestial bodies including galaxies pairs, double black holes etc. 
The obvious characteristics of double peaks generally exist on emission lines, which are composed of two or more peaks within specific wavelength ranges.

Finding spectra with double-peaked emission lines is different due to several reasons, as follows:
First, high dimensionality is included in a spectrum. The dimension number reaches up to 4000 in each spectrum, leading to the difficultly of searching for rare double peaks.
Second, the dimensions showing double peaks are very small relative to the total spectra. Thus, the search for double peaks from high-dimensional data is difficult when using conventional methods.
Finally, in a double-peaked emission lines spectrum, the double peaks always exhibit different profiles.

Based on the above characteristics of data with double-peaked emission lines, data preprocessing is worked out before experiments. A characteristic extraction method based on relevant subspace (RS) is used to obtain the characteristics related with double peaks\cite{qu2019dops,cai2019feature,zhang2016relevant}. A dataset including 345 spectra with double-peaked emission lines, which are known and identified currently, is selected from LAMOST DR5 to extract the useful characteristics\cite{wang2018double,shi2014search}. 

First, a new dataset is obtained by data normalization and wavelength intercept. Second, the local dataset $LDS(o, d)$ of object $o$ on dimension $d$ is computed by KNN (K-nearest neighbour) algorithm. Third, a global density matrix is built by using the local density of each object on all dimensions. Fourth, the difference between dimensions is calculated by the density matrix to measure the difference between attributes. Finally, the attributes with higher difference are extracted as the members of the relevant subspace. A characteristic subspace of length 8 is obtained by analysing the relevant subspace.
Eight emission lines are considered as the characteristics likely to appear as double peaks, including H$\alpha$, [OIII]$\lambda\lambda$4959,5007, H$\beta$, [NII]$\lambda\lambda$6548,6584, [SII]$\lambda\lambda$6717,6731. The dimensions of each line are listed in Tabel \ref{tab_lines}, including the beginning and ending values of wavelength.

\begin{table}[htb]
\centering
\setlength{\abovecaptionskip}{0pt}
\setlength{\belowcaptionskip}{10pt}
\setlength{\tabcolsep}{4mm}
\caption{Formal Context of Double-peaked Emission Lines Spectra}
\label{tab_lines}
\begin{tabular}{ccc}
\hline
Line&begin wave/(\AA)&end wave(\AA)\\
\hline
$H\beta$&4859&4866\\
$[OIII]\lambda$4959&4956&4963\\
$[OIII]\lambda$5007&5002&5012\\
$[NII]\lambda$6548&6542&6559\\
$H\alpha$&6555&6573\\
$[NII]\lambda$6584&6577&6594\\
$[SII]\lambda$6717&6717&6726\\
$[SII]\lambda$6731&6721&6737\\
\hline
\end{tabular}
\end{table}

Data preprocessing of relevant subspace is a meaningful process that reduces the dimensions of data and time complexity. The dimensional disaster in massive and high-dimensional data can be effectively avoided. Meanwhile, the most useful information for searching of double peaks is extracted from all dimensions, which centralizes the data information. The dataset after characteristic extraction is used to the build the SVM-Lattice for double-peaked emission liens.

\subsection{SVM-Lattice Construction}
Among the 345 spectra with double-peaked emission lines, 341 spectra include all 8 characteristics, while the remaining spectra are shown without the last 2 characteristics due to a larger redshift. To ensure that all characteristics are contained in the spectra, 341 spectra with all features are applied in the construction of the SVM-Lattice.

\subsubsection{Initial Lattice}
A formal context containing characteristics and objects must be provided before constructing a formal concept lattice. In this paper, a double-peaked emission lines sample including 341 objects is selected as a dataset of formal context. The characteristics in the formal context are H$\alpha$, [OIII]$\lambda\lambda$4959,5007, H$\beta$, [NII]$\lambda\lambda$6548,6584, [SII]$\lambda\lambda$6717,6731. For each characteristic, 341 spectra are labelled according to the existence of objects with these characteristics. The formal context of our spectra data is shown in Table \ref{tab2}, composed of values 1 and 0. The first column is the index of samples with double-peaked emission lines from 1 to 341. The rest of the columns are labels of objects on each characteristic. The value of 1 denotes that the double peaks exists for the $ith$ object with respect to the $jth$ characteristic, while 0 indicates the opposite.

\begin{table*}[htp]
\centering
\setlength{\abovecaptionskip}{0pt}
\setlength{\belowcaptionskip}{10pt}
\caption{Formal Context of Double-peaked Emission Lines Spectra}
\label{tab2}
\begin{tabular}{ccccccccc}
\hline
Object&F1(H$\beta$)&F2([OIII]$\lambda4959$)&F3([OIII]$\lambda5007$)&F4([NII]$\lambda 6548$)&F5(H$\alpha$)&F6([NII]$\lambda 6584$)&F7([SII]$\lambda 6717$)&F8([SII]$\lambda 6731$)\\
\hline
O1&0&1&1&1&1&1&1&1\\
O2&1&1&0&0&0&0&0&0\\
O3&1&1&1&0&0&0&1&1\\
O4&1&1&0&0&1&1&1&1\\
O5&1&0&0&1&1&0&0&1\\
O6&1&1&1&1&1&1&1&1\\
O7&1&1&1&1&1&1&1&0\\
O8&1&1&1&0&1&1&1&0\\
O9&1&1&1&0&0&0&1&1\\
O10&0&0&1&0&0&0&1&1\\
O11&1&0&0&0&1&1&1&1\\
O12&0&0&0&1&0&1&1&1\\
$\vdots$& & & & & & & & \\
O341&1&1&1&0&1&1&1&0\\
\hline
\end{tabular}
\end{table*}

\begin{figure*}[htb]
\centering
\includegraphics[scale=0.5]{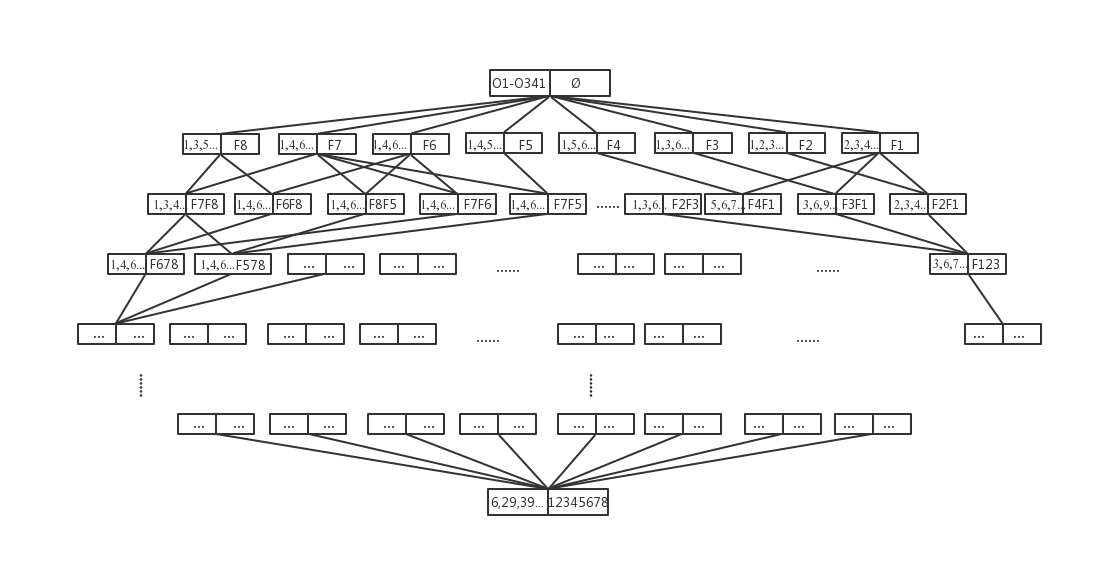}
\caption{An Expendable Legend of the Completed Lattice}
\label{fig7}
\end{figure*}

According to the formal concept in Table \ref{tab2}, an initial concept lattice $L$ including 256 nodes can be built. The intent of the root node on the initial lattice is null; meanwhile extent is all of the samples. The last node is connection of all characteristics and objects on these characteristics. 
In the initial lattice, 8 layers exist, with several numbers of nodes according to the formal concept lattice. The complete drawing of the initial lattice is shown in Figure \ref{fig7}. It is observed that the nodes in the $ith$ layer represent a subsection of $i-1$ nodes in the upper layer. On the same layer, the number of characteristic combination of each node is the same. In Figure \ref{fig7}, the nodes with single intent are on the second layer. Below the third floor, at least two characteristics are contained in the nodes in each layer. There are 9 layers on the diagram, in which the node numbers on the first and last layers are 1. The node in the first layer is a combination of all objects with null characteristic, while several objects and all characteristics are contained in the last last.

\begin{figure*}
\setlength{\abovecaptionskip}{0.cm}
\setlength{\belowcaptionskip}{-0.cm}
\centering
\includegraphics[scale=0.5]{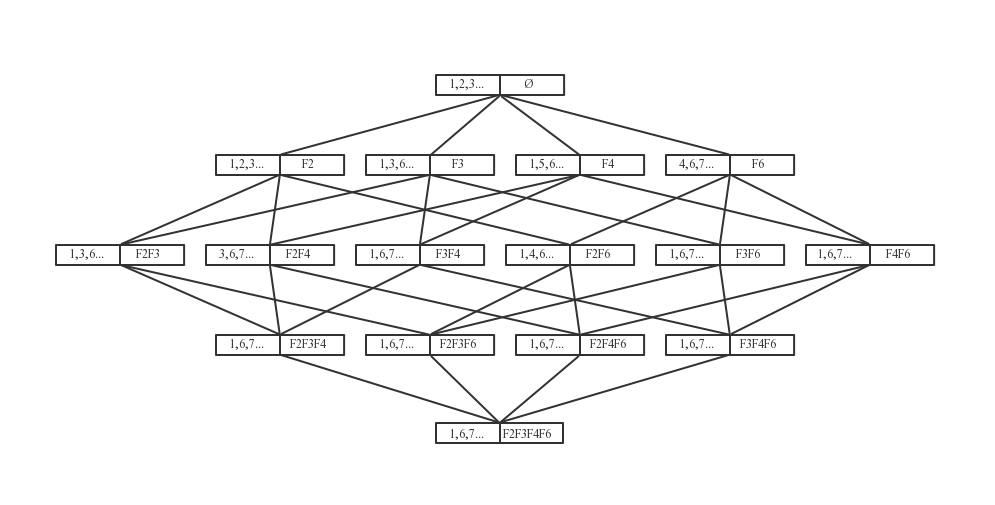}
\caption{An Example Lattice for 341 Objects on 4 Features}
\label{fig8}
\end{figure*}

For a more intuitive understanding of the initial lattice, several characteristics are selected to serve as a specific example, as shown in Figure \ref{fig8}. A subsample including 341 objects on F2[OIII]$\lambda4959$), F3([OIII]$\lambda5007$), F4([NII]$\lambda 6548$), F6([NII]$\lambda 6584$) is used to construct a sublattice as an sample. There are 5 layers in Figure \ref{fig8}, composed of 16 nodes in total. One characteristic is represented by each node on the second layer, containing 4 nodes. The nodes with the same number are in the fourth layer, with 3 characteristics in combination. The layer with the most nodes is the third, for which there are subsections of each of the two nodes on the last layer. In addition, the nodes on the last two layers are connected by 3 and 4 nodes, respectively. The example can serve as a reference for the total lattice.

\subsubsection{Completed SVM-Lattice}
The extents in the initial lattice are from formal context and must be mapped by the DoPS. In each node, the extents are replaced by positive sample from SVM as new extent. The training dataset in the training process of SVM is determined by the positive sample with respect to characteristics according to the formal context. For one characteristic combination, the numbers of positive and negative samples are the same for training and testing datasets. A hyperplane is obtained from the training dataset using the DoPS. The positive sample in the testing dataset upper hyperplane is regarded as the new extent of the node. DoPS is used as a mapping method for the SVM-Lattice, which is a completed lattice.

The mapping process of each node $h(o, f)$ in the initial lattice is as follows. Extent $o$ is regarded as the positive training dataset $PD$ on characteristic $f$. Select negative sample $ND$ without double-peaked emission lines on characteristic $f$; combine two-thirds of $PD$ and $ND$ to be training dataset $TR$ on characteristic $f$, meanwhile, the remaining third serves as the testing dataset $TE$ on characteristic $f$; train a classifier (hyperplane) for double-peaked emission lines data using $TR$; record the support vectors and the testing sample upper hyperplane, named $SV$ and $PT$ on $f$, respectively; obtain the accuracy and recall rate by $TE$ to serve as reference values; construct a hyperplane concept including $SV$ and $PT$ as the intent and extent of $h$ respectively; and add the new node $h'(PT, SV)$ to the SVM-Lattice with the accuracy and recall rate.

A new lattice named SVM-Lattice is constructed by the DoPS method, which is the main work in the construction process. For each node in the SVM-Lattice, the accuracy and recall rate are recorded as additional information regarding evaluation of the classification result on characteristic combinations of this node. All node information is listed in Table \ref{tab3}, including the intent, extent, accuracy, recall rate, and shapes of training and testing datasets. In the construction of the concept lattice, the intent is the common characteristic set owned by extents. Thus, in the first node, the intent is empty due to the extent including all samples of formal context.
\begin{table*}[htp]
\centering
\setlength{\abovecaptionskip}{5pt}
\setlength{\belowcaptionskip}{10pt}
\caption{The Information of Nodes in the SVM-Lattice}
\setlength{\tabcolsep}{6mm}
\label{tab3}
\begin{tabular}{ccccccc}
\hline
Node&Intent&Extent&Training Shape&Testing Shape&Recall Rate&Accuracy\\
\hline
1&Null& & & & & \\
2&F2F3F4F5F6F7F8&1,2,3,4...&43$\times$69&15$\times$69&1&0.4\\
3&F2&1,2,3,4...&454$\times$7&152$\times$7&1&0.5\\
4&F1F2&1,2,3,4...&198$\times$14&67$\times$14&1&0.402\\
5&F2F3F7F8&1,2,3,4...&89$\times$33&45$\times$33&1&0.41\\
6&F1F2F3F7F8&1,2,3,4...&80$\times$40&27$\times$40&1&0.407\\
7&F2F5F6F7F8&1,2,3,4...&88$\times$48&30$\times$48&1&0.4\\
8&F2F7F8&1,2,3,4...&171$\times$24&58$\times$24&1&0.396\\
9&F1F2F7F8&1,2,3,4...&108$\times$31&37$\times$31&1&0.405\\
10&F1F2F5F6F7F8&1,2,3,4...&61$\times$55&21$\times$55&1&0.380\\
11&F4F5F8&1,2,3,4...&116$\times$35&40$\times$35&1&0.4\\
12&F1&1,2,3,4...&454$\times$7&152$\times$7&1&0.5\\
$\vdots$\\
256&F1F3F4F5F7F8&1,2,3,4...&60$\times$57&20$\times$57&1&0.4\\
\hline
\end{tabular}
\end{table*}

\subsubsection{Reduced Lattice}
According to the formal context in Table \ref{tab2}, all characteristic combinations are considered to be intents in the completed SVM-Lattice. However, users are attracted by parts of the lattice, instead of the complete lattice. For enhanced convenience for users, the pruning process is considered to reduce the completed SVM-Lattice. The association rules method finds meaningful correlation rules with specific coefficient between objects from massive data. To obtain a reduced SVM-Lattice, the Apriori method is applied to remove the characteristic combinations with lower coefficients. After setting the support threshold $s$, the mining of frequent items is premise of rule finding. A rule of the form $X\Rightarrow Y$ is analysed according to its coefficient $c$ for determining the useful characteristic combinations.

In the pruning process, the most frequent items are determined by the set of support threshold $s$. Therefore, the value $c$ must be set before finding frequent items. In addition, the coefficient threshold $c$ must be determined manually when the removed characteristics are selected. In this paper, $s$ and $c$ are set as 0.2 and 0.5, respectively, according to background knowledge. The nodes removed from the SVM-Lattice by pruning are listed in the Table \ref{tab4}.

\begin{table}
\centering
\setlength{\abovecaptionskip}{5pt}
\setlength{\belowcaptionskip}{10pt}
\caption{Pruning Nodes in SVM-Lattice}
\label{tab4}
\setlength{\tabcolsep}{7mm}
\begin{tabular}{ccc}
\hline
Node&Intent&Extent\\
\hline
126&F2F3F5&1,6,7...\\
268&F2F6F8&1,4,6...\\
110&F2F4F8&1,6,17...\\
95&F2F4F7&1,6,7...\\
36&F1F7F8&3,4,6...\\
46&F4F6F7&1,6,7...\\
20&F2F3F7&1,3,6...\\
37&F5F6F7F8&1,4,6...\\
112&F1F2F3&3,6,7...\\
$\vdots$\\
95&F2F4F7&1,6,7...\\
\hline
\end{tabular}
\end{table}

There are 66 nodes included in Table \ref{tab4}, that are not relatively meaningful for users. The first column in this table represents the numbers of removed nodes in the SVM-Lattice. The last two columns are the characteristic combinations and the positive sample upper hyperplane in testing dataset, respectively. The new lattice is named reduced SVM-Lattice, for which a final lattice has been pruned. The pruning process is also worked when a completed SVM-Lattice is built during pruning, instead of after building the lattice. The time cost is reduced when pruning occurs during the build of the SVM-Lattice instead of outside of construction.

\subsection{Evaluation Result}
The SVM-Lattice is used as a classifier for data with double-peaked emission lines, meanwhile, the confidence of each node is also proposed. First, we give a classification result of the SVM-Lattice based on a dataset of size 10000. The best node for each object is found by comparing the nodes in the SVM-Lattice. The index of nodes for several objects is listed in Table \ref{tab_classi}. The confidence of each classification result is given according to the basic information of the SVM-Lattice. 

The best node close to each point is found by iterating through all nodes in the SVM-Lattice. Table \ref{tab_classi} shows the best node number and the corresponding confidence for each object. In fact, the best node just is simply the node with the smallest distance to the object. The point may be between two nodes due to the natures of the SVM-Lattice. To offer an intuitive perspective, the node closest to each point is proposed in this table. In the SVM-Lattice, each node is shown with the corresponding confidence by the association rules algorithm. Confidence is a judgement value of the credibility of a classification result. The higher the confidence value is, the more reliable the result is.

\begin{table}
\centering
\setlength{\abovecaptionskip}{5pt}
\setlength{\belowcaptionskip}{10pt}
\caption{Classification Result of Double Peaks}
\label{tab_classi}
\setlength{\tabcolsep}{5mm}
\begin{tabular}{ccc}
\hline
Object Index&Best Node&Confidence\\
\hline
1&26&0.266\\
2&176&0.472\\
3&26&0.266\\
4&26&0.266\\
5&26&0.266\\
6&32&0.166\\
7&32&0.166\\
8&26&0.266\\
9&109&0.337\\
10&68&0.217\\
11&12&0.686\\
$\vdots$\\
10000&26&0.266\\
\hline
\end{tabular}
\end{table}

\subsection{An Example for Specific Features}
\begin{figure}[htp]
\centering
\includegraphics[scale=0.35]{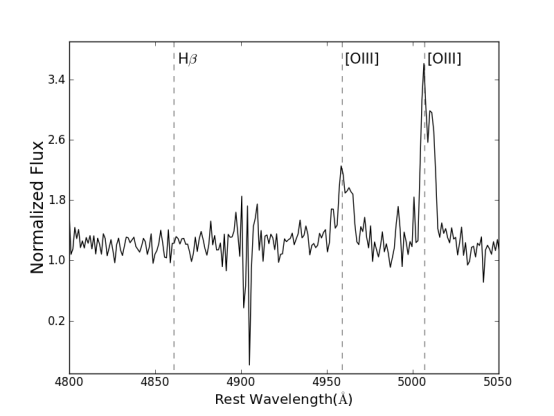}
\caption{An Example of Double-Peaked Emission Lines}
\label{fig_double}
\end{figure}

The profiles of double peaks are composed of at least two peaks in a specific location in spectra. A spectrum with double-peaked emission lines is selected as an example, as shown in Figure \ref{fig_double}. Part of a total spectrum is drawn in Figure \ref{fig_double}, in which the horizontal and vertical coordinates denote the wavelength and normalized flux of a spectrum, respectively. The obvious double peaks exist on $[OIII]\lambda4959$ and $[OIII]\lambda5007$, while this characteristic is weak on H$\beta$. A sample of classification for the three characteristics is given to show the results obtained using the SVM-Lattice.

\begin{figure}[htp]
\centering
\includegraphics[scale=0.5]{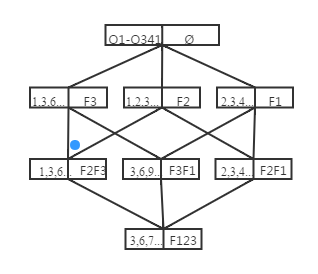}
\caption{Classification Result of F1FF3. The blue point is denoted the final location of classification for the spectrum with strong double peaks on F2 and F3.}
\label{fig_f1f2f3}
\end{figure}

Before classification of the spectrum, a small SVM-Lattice is built using a sub-dataset with H$\beta$(F1), $[OIII]\lambda4959$(F2) and $[OIII]\lambda5007$(F3). The location of the object in new SVM-Lattice is shown in Figure \ref{fig_f1f2f3}. The object matches node in the SVM-Lattice according to the recognition and evaluation process. The distance between object and node is calculated to be a measurement of the matching result. The $dis(o,F)$ denoted the distance between object $o$ and node with $F$. The $dis(o,F3)$ is 2.5, which is smaller than $dis(o,F2)$ with 2.8 and larger than $dis(o,F2F3)$ with 1.2. Thus, the object is located between the nodes with $F3$ and $F2F3$. The node with $F2F3$ have the confidence 0.8, which denotes the credibility of classification on this node.

\subsection{Result Discussion}
To test and verify the usability of the proposed method, multiple sets of data sample are used in our experiment. 
A sample is selected randomly from LAMOST to test the effectiveness of our SVM-Lattice compared with the template matching method. In this paper, the spectra with double-peaked emission lines are the main testing data for the experiments. For each characteristic combination, the positive sample with double-peaked emission lines is regarded as a template to be matched with the testing sample. The distance between one object in testing sample and positive sample are separately computed to choose a minimum value as the matching result. A given object in the testing sample is compared with all template data for each characteristic combination. The testing data are assigned to a class in which the matching template is nearest to the classed object.

Compared with template matching, the SVM-Lattice presents some advantages that are absent in the last method. The SVM-Lattice can find the best intent on one certain layer to provide an accurate classification of given object. Moreover, the distance between layers can be computed according to the SVM-Lattice. In addition to the distance between layers, the distances between nodes on the same layer are obtained from the SVM-Lattice to be compared with others. 
The distances can be obtained from the SVM-Lattice, but not from template matching, are listed in Table \ref{tabcompare}.

\begin{table}
\centering
\setlength{\abovecaptionskip}{5pt}
\setlength{\belowcaptionskip}{10pt}
\caption{Comparison of Advantages Between Two Methods}
\label{tabcompare}
\setlength{\tabcolsep}{2mm}
\begin{tabular}{ccc}
\hline
&SVM-Lattice&Template matching\\
\hline
Accuracy&$\surd$&$\times$\\
Layer&$\surd$&$\times$\\
Distance between layers&$\surd$&$\times$\\
Distance between nodes&$\surd$&$\times$\\
Distance between objects&$\surd$&$\surd$\\
Short circuit in search&$\surd$&$\times$\\
Less time consumption&$\surd$&$\times$\\
\hline
\end{tabular}
\end{table}

It can be observed that the first four distances can be obtained based on the layers in the SVM-Lattice. The distance between the objects according to template matching is computed by appending the single object instead of data combinations. However, in the SVM-Lattice, it is determined according to the distance between the object and hyperplane, which is different from the previous result. The search process will be interrupted when the smallest distance occurs. It is not necessary to continue searching after the best result is found. Thus, the short circuit exists in the SVM-Lattice, aiming to reduce the search cost. The last line in Table \ref{tabcompare} shows the superior performance of the SVM-Lattice versus template matching, which is due to the reduced time complexity.

The spectra from LAMOST for each characteristic combination are selected randomly to be matched with the model template including 341 positive objects. In the matching process, Gaussian fitting is used to transform the object to a Gaussian profile due to the specifics of double-peaked emission lines. 
For every node, each point of the testing dataset is fitted by a Gaussian function to be matched with one fitted from model data. The two Gaussian functions are computed to obtain similarity between unknown data and template. The best node is found, for which the testing data are similar with respect to the template. On the best node, the testing data are most similar with the template compared with those of other nodes. 

The testing data are computed to determine the nearest node according to the distance with respect to negative and positive templates. The steps of the matching method for the $ith$ testing object are as follows: obtain Gaussian functions $f(x_i)$ and $g(x_j)$ fitted by $ith$ tested object and $jth$ template on each node; compute the distance between $ith$ tested object and all templates by $f(x)-g(x)$, which are named $Dis(i)$; find the template nearest to the $ith$ object on each node; compose the smallest distance between $ith$ and template on all nodes; and select the best intent that has the greatest similarity between the testing object and template.

All predictions from DoPS of the testing dataset on different nodes are compared with each other to find the closest hyperplane for one intent. To find an optimized node for a given object from the SVM-Lattice, the nodes with different intents are computed from top to bottom. For a detail comparison process, the lattice in Figure \ref{fig8} is shown as an example. The intent of F2F3 on the third layer is the intersection of two intents with F2 and F3. The more that are characteristics included in a node, the more useful the information is in the data, and the greater the credibility is of the classification result. Each node is denoted as a unique hyperplane by DoPS, which represents a measurement factor for a given object. The similarity calculation of distance between the object and different hyperplanes is initiated from the second layer. The smaller the similarity is between one hyperplane and object, the more possible it is for the object to the applied in the intent.

The results of finding the best node obtained by the SVM-Lattice and model matching method are listed in Table \ref{tab5}, including the similarity between template matching and prediction according to the SVM-Lattice of each testing object.

\begin{table*}[htp]
\centering
\setlength{\abovecaptionskip}{5pt}
\setlength{\belowcaptionskip}{10pt}
\caption{Finding Result of Model Matching Method}
\label{tab5}
\setlength{\tabcolsep}{6mm}
\begin{tabular}{ccccc}
\hline
testing object&predict&similarity&best node of SVM-Lattice&best node of template matching\\
\hline
1&-4.246&0.582&12&1\\
2&-0.831&0.444&76&1\\
3&-3.506&0.724&226&11\\
4&-4.617&0.399&226&1\\
5&-4.099&0.593&226&34\\
6&-2.041&0.432&107&34\\
7&-1.588&0.427&179&34\\
8&-4.194&0.440&226&34\\
9&-4.021&0.667&226&10\\
10&-2.674&0.682&126&1\\
11&-1.804&0.389&126&34\\
$\vdots$\\
100&-0.983&1.184&46&34\\
\hline
\end{tabular}
\end{table*}

The 30 objects in the testing data are selected to find the best node among 256 nodes due to the large dataset. The negative predictions are all included in Table \ref{tab5} due to the prediction function obtained from DoPS. The predictions are regarded as the similarity between testing object and hyperplane. The smaller the predicted value is, the closer the distance is between the hyperplane and objects. The best node is found by comparing all distances to obtain the minimum distance. Thus, the intent with the smallest distance is considered the best node of one object existing in Table \ref{tab5}. 

\subsection{Effectiveness Testing}
\begin{figure}
\centering
\includegraphics[scale=0.4]{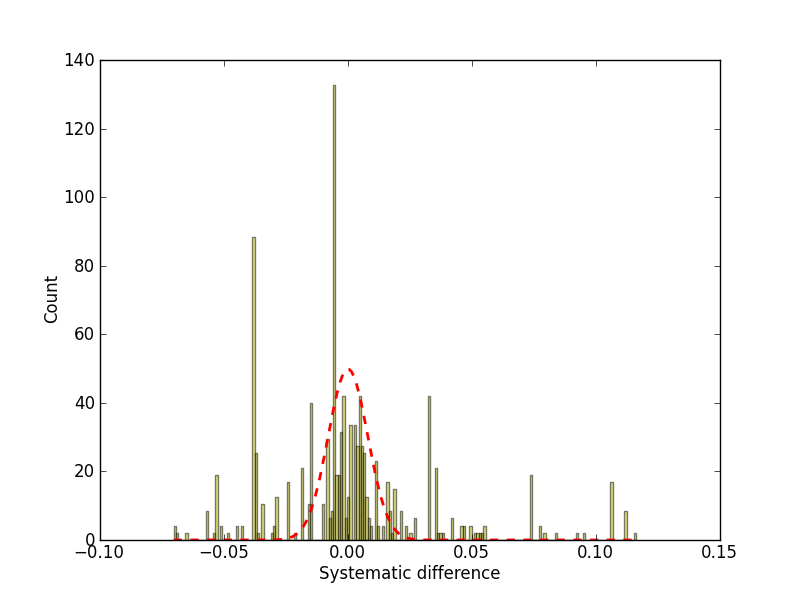}
\caption{Comparison of Consistence}
\label{fig_hist}
\end{figure}

To test the effectiveness of the SVM-Lattice, the subtraction of SVM-Lattice and template matching is obtained to observe the consistency of the two methods. The function with $y=o(p)-o(s)$ is computed as a measurement of the consistency, where $o(p)$ and $o(s)$ are the closed objects found on each node using SVM-Lattice and template matching, respectively. The closer to 0 $y$ is, the higher the consistency is. The dataset including 10000 objects is selected to obtain a histogram, which is drawn in Figure \ref{fig_hist}. 

In Figure \ref{fig_hist}, it is observed that the histograms near 0 are denser. Gaussian fitting is used to fit the trend of the denser areas. The histogram with 0 is the highest, and the others are closer to 0. In total, it can be concluded that our SVM-Lattice is effective by comparing with template matching.

\begin{figure}
\centering
\includegraphics[scale=0.4]{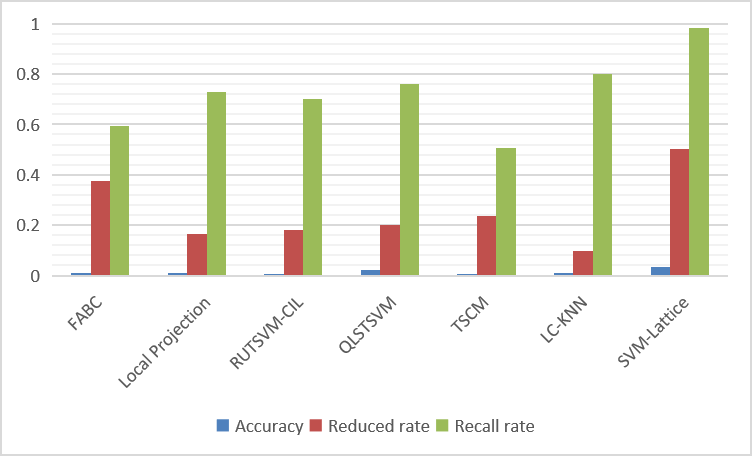}
\caption{Comparison of SVM-Lattice and Other Algorithms}
\label{rate}
\end{figure}

To verify the effectiveness of the SVM-Lattice, four methods are used to demonstrate the stability, including FABC\cite{ezghari2017new}, Local projection\cite{ortner2017local}, TSCM\cite{zelenkov2017two}, LC-KNN\cite{deng2016efficient}, RUTSVM-CIL\cite{B.R2020a} and QLSTSVM\cite{qian2019qua}. The dataset sized 10000 from LAMOST DR5 is selected to obtain the accuracy, recall and reduced rate of methods, which is shown in the Figure \ref{rate}. It is seen that the recall and reduced rate of SVM-Lattice are higher than others. Meanwhile, the accuracy of SVM-Lattice is also better than other methods with less than 0.1. Overall, the SVM-Lattice shows better performance form the figure, which tests the effectiveness of the SVM-Lattice.

\subsection{Efficiency Testing}
\begin{figure*}
\centering
\subfigure[50 nodes]{
\includegraphics[scale=0.42]{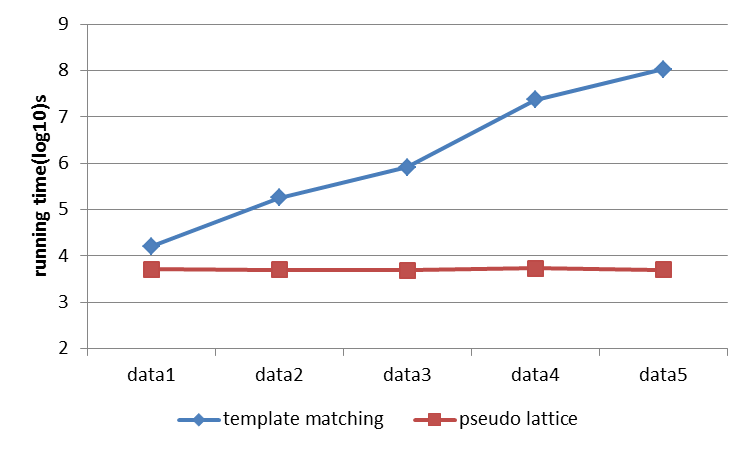}
\label{time_a}}
\subfigure[100 nodes]{
\includegraphics[scale=0.42]{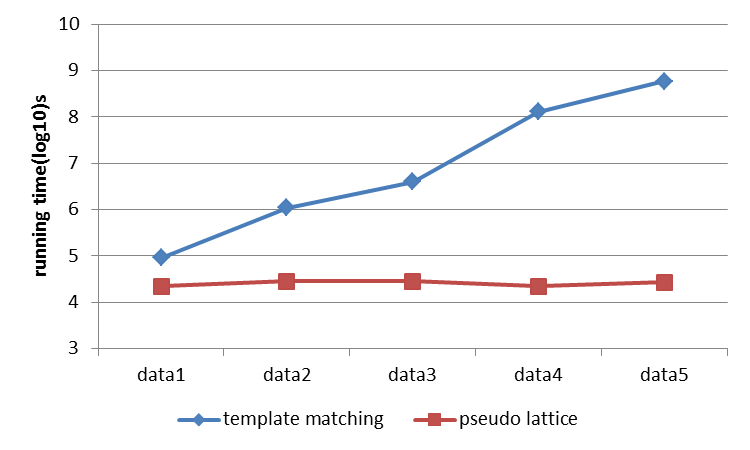}
\label{time_b}}
\subfigure[all nodes]{
\includegraphics[scale=0.42]{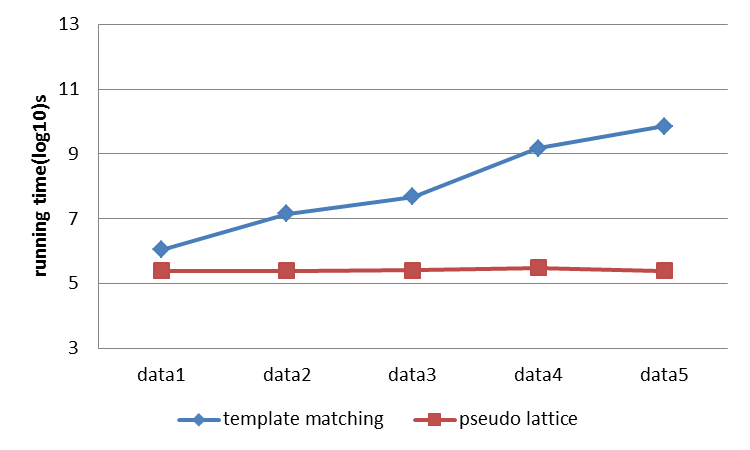}
\label{time_c}}
\caption{Running Time of Two Methods}
\label{time}
\end{figure*}

The SVM-Lattice in this paper is an evaluation method for special data with characteristics. To test the efficiency of this proposed method, five datasets with different sizes on three characteristic combinations are used for analysis of running time. Datasets sized 100, 500, 1000, 5000, 10000, from LAMOST DR5, are applied to verify the efficiency of the SVM-Lattice. Compared with template matching, the SVM-Lattice is built based on the DoPS, which is ahead of the prediction of testing data.
The running time of two methods, shown in Figure \ref{time}, are quantified with respect to intents for different sizes of datasets. Data1, data2, data3, data4 and data5 denote five testing datasets including 100, 500, 1000, 5000, and 10000 objects, respectively.

Figure \ref{time} shows the running times of two methods with five datasets for three characteristic combinations. The three sub-figures \ref{time_a}, \ref{time_b}, and \ref{time_c} exhibits intents including 50, 100, and 256 nodes, respectively. It is seen that the running time of the SVM-Lattice is always less than that of template matching; meanwhile, the time cost is stable regardless of the dataset size. For different number of nodes, the SVM-Lattice exhibits better performance on various testing datasets than template matching does.

\section{Summary}
\label{Summary}
In this paper, we propose a novel method, SVM-Lattice, which is based on the DoPS and formal concept lattice, to perform a systematic evaluation of the classification results for special data with rare characteristics. The main works pertaining to the method in this paper are as follows: first, the definition of the new hyperplane concept as mapped based on the DoPS is given, including the intent and extent of each node generalized and specialized by relationship between layers; second, the SVM-Lattice building algorithm, pruning algorithm based on the association rules and evaluation algorithm are proposed to complete the total method; and finally, the double-peaked emission lines spectra data are used to build a SVM-Lattice as an example of special data with characteristics. The effectiveness and efficiency of our method are proven on five datasets with different sizes of various intents. Eight characteristics are included in the formal context, which contains 341 objects with double-peaked emission lines. An evaluation of the accuracy and recall rate of the classification for double peaks is given for different characteristic combinations, which can serve as an evaluation of DoPS classification from several characteristic angles.

\section*{Acknowledgement}
\label{Acknowledgement}
This study is supported by the National Natural Science Foundation of China (Grant Nos. U1931209, U1731126) and Shanxi Province Key Research and Development Program (Grant Nos. 201903D121116, 201803D121059)

The Guo Shou Jing Telescope (the Large Sky Area Multi Object Fiber Spectroscopic Telescope, LAMOST) is a National Major Scientific Project built by the Chinese Academy of Sciences. Funding for the project has been provided by the National Development and Reform Commission. LAMOST is operated and managed by National Astronomical Observations, Chinese Academy of Sciences.

\bibliographystyle{ieeetr}
\bibliography{bibfile}

\EOD

\end{document}